\documentstyle[bezier,12pt]{article}
\renewcommand{\thefootnote}{\fnsymbol{footnote}}

\newcommand{\newsection}{    
\setcounter{equation}{0}
\section}
\def\appendix#1{
  \addtocounter{section}{1}
  \setcounter{equation}{0}
  \renewcommand{\thesection}{\Alph{section}}
  \section*{Appendix \thesection\protect\indent \parbox[t]{11.715cm} {#1} }
  \addcontentsline{toc}{section}{Appendix \thesection\ \ \ #1}
  }
\newcommand{\tr}[1]{\,{\rm tr}\,#1}
\newcommand{\ntr}[1]{\,\frac1N {\rm tr}\,#1}
\def\e{{\,\rm e}\,}

\def\eop{\vspace*{\fill}\pagebreak}
\def\be{\begin{equation}}
\def\ee{\end{equation}}
\def\bea{\begin{eqnarray}}
\def\eea{\end{eqnarray}}
\def\LB{\left(}
\def\RB{\right)}
\def\LA{\left\langle}
\def\RA{\right\rangle}
\def\RAG{\right\rangle_{\hbox{\footnotesize{Gauss}}}}
\def\RAH{\right\rangle_{\hbox{\footnotesize{Haar measure}}}}
\newcommand{\rf}[1]{(\ref{#1})}
\newcommand{\eq}[1]{Eq.~(\ref{#1})}
\newcommand{\bbox}[1]{\mbox{\boldmath {$\displaystyle #1$}}}

\def\a{\alpha}

\def\l{\lambda}

\newcommand{\non}{\nonumber \\*}

\def\u{\mbox{\boldmath $u$}}
\def\ud{{\mbox{\boldmath $u$}}^\dagger}
\def\j{\mbox{\boldmath $j$}}
\def\lbr{{\bf \Big\{}}
\def\rbr{{\bf \Big\}}}
\newcommand{\ie}{{\it i.e.}\ }

\newcommand{\ra}{\rightarrow}
\newcommand{\fr}[2]{{\textstyle {#1 \over #2}}}

\begin{document}

\begin{titlepage}
\begin{flushright}
ITEP-TH-13/95 \\
December, 1995 \\
\small{hep-th/9601139}
\end{flushright}
\vspace*{.1cm}

\begin{center}
{\LARGE Supersymmetric Matrix Models \\[.3cm] and the Meander Problem}
\end{center} \vspace{.3cm}

\begin{center}
{\large Yuri Makeenko}\footnote{~E--mail:
\ makeenko@vxitep.itep.ru \ / \
 makeenko@nbi.dk \ } \\
\vskip 0.2 cm
{\it Institute of Theoretical and Experimental Physics,}
\\ {\it B. Cheremushkinskaya 25, 117259 Moscow, Russia}
\\ \vskip .1 cm
and  \\  \vskip .1 cm
{\it The Niels Bohr Institute,} \\
{\it Blegdamsvej 17, 2100 Copenhagen, Denmark} \\
\vspace{0.2cm} \mbox{} \\ {\large and} \\
\vspace{0.2cm} \mbox{} \\
{\large Iouri Chepelev}  \\ \vskip 0.2 cm
{\it Department of Physics, Moscow State University,} \\
{\it Vorobiyevie Gori, 119899 Moscow, Russia}
\end{center}

\vskip 0.5 cm
\begin{abstract}
We consider matrix-model representations of the
meander problem which describes, in particular, combinatorics for
foldings of closed polymer chains.
We introduce a supersymmetric matrix model for describing
the principal meander numbers. This model is of the type
proposed by Marinari and Parisi for discretizing a
superstring in $D=1$ while the supersymmetry is
realized in $D=0$ as a rotational symmetry between bosonic
and fermionic matrices.
Using non-commutative sources, we reformulate
the meander problem in a Boltzmannian
Fock space whose annihilation and creation operators
obey the Cuntz algebra.
We discuss also the relation between the matrix models
describing the meander problem and the Kazakov--Migdal model
on a $D$-dimensional lattice.
\end{abstract}

\eop
\end{titlepage}

\setcounter{page}{2}
\renewcommand{\thefootnote}{\arabic{footnote}}
\setcounter{footnote}{0}

\newsection{Introduction}

Matrix models is a standard tool for describing
discretized random surfaces (or, equivalently, strings)~\cite{Kaz85}.
A supersymmetric extension of this construction
was first proposed by Marinari and Parisi~\cite{MP90}
and studied for the $D=1$ dimensional target space.

We introduce in the present paper supersymmetric matrix
models in the $D=0$ dimensional target space which differ
from the Hermitean supermatrix models of Ref.~\cite{AG91} and
are a version of the Marinari--Parisi construction
in $D=0$. Our model deals with the ``superfields''
\be
W_a= \left(B, F\right),~~~~~~\bar{W}_a= \left(B^\dagger, \bar{F}\right)
\label{superfields}
\ee
where $a=1,2$ while $B$ and $F$ are complex bosonic and fermionic
(\ie Grassmann valued) $N\times N$ matrices, respectively.

Since the propagators for both bosonic and fermionic matrices coincide:
\bea
\LA  B_{ij} B^\dagger_{kl}\RAG &=& \frac 1N \delta_{il} \delta_{kj} \,, \non
\LA  F_{ij} \bar{F}_{kl}\RAG &=& \frac 1N \delta_{il} \delta_{kj} \,,
\label{BFpropagators}
\eea
the supersymmetry reduces in $D=0$ simply to rotations between the
$B$- and $F$-compo\-nents. The proper transformation reads
\bea
\delta B = \bar\epsilon F\,,& &~~~\delta F = -\epsilon B\,, \non
\delta B^\dagger =\bar F \epsilon\,,
& &~~~\delta \bar F = - B^\dagger \bar\epsilon\,,
\label{huge}
\eea
where $\epsilon$ and  $\bar\epsilon$ are Grassmann valued.

Any potential, which is symmetrically
constructed from the ``superfields''~\rf{superfields}, is supersymmetric
so that contributions from the loops of the bosonic and fermionic matrix
fields are mutually cancelled which is the key property of the supersymmetry.
The simplest Gaussian supersymmetric potential reads
\be
V_{\rm Gauss}=N \sum_{a=1}^2 \tr \bar{W}_a W_a \equiv
N \tr \left( B^\dagger B +\bar{F} F\right),
\label{SVGauss}
\ee
which reproduces the propagators~\rf{BFpropagators}.
It is obviously invariant under the rotation~\rf{huge} even when
$\epsilon$ and $\bar\epsilon$ are fermionic $N\times N$ matrices.
It is also clear from \eq{SVGauss} why one needs complex matrices
in $D=0$: the trace of the square of a fermionic matrix vanishes.

We elaborate in this paper the technique for dealing with the
supersymmetric matrix models on an example of the one
which describes combinatorics of the meander numbers.
This challenging combinatorial problem, which is
described in Sect.~2, is not yet solved.
The associated matrix model, which describes a physical problem
of enumerating different ways of foldings of a closed polymer chain,
is of a next level of complexity with respect to the Hermitean
one- or two-matrix models, the multi-matrix chain and
its multi-dimensional extension --- the Kazakov--Migdal model~\cite{KM92}.

We introduce in Sect.~2 the complex matrix model which is equivalent
to the Hermitean one~\cite{KK} in describing the meander numbers.
We construct then the supersymmetric matrix model which
describes the principal meanders.
We discuss also the relation between the matrix models
describing the meander problem and the Kazakov--Migdal model
on a $D$-dimensional lattice.

Using non-commutative sources, we reformulate in Sect.~3
the meander problem as a problem of averaging in a Boltzmannian
Fock space whose annihilation and creation operators
obey the Cuntz algebra. The averaging expression is represented
in the form of a product of two continued fractions.

The Appendix~A contains a solution of the combinatorial problem of
summing over words built up of unitary matrices, which is
equivalent to the Kazakov--Migdal model
with the Gaussian potential, via free random variables.

In the Appendix~B we demonstrate how the equations of Sect.~3,
which are obtained using the matrix-model representations, can be 
alternatively derived pure combinatorially.

We comment in the Appendix~C  on a possibility of solving
the meander problem via free random variables.
We show that this approach does not work for the meander problem
since the variables are not free for this case
so that the theorem of addition of free random variables is not applicable.

\newsection{Matrix models for the meander problem}

The meander problem is known to people working on Quantum Field Theory
since the middle of the eighties from V.~Arnold.
The problem is to calculate combinatorial numbers associated with
the crossings of an infinite river (Meander) and a closed road
by $2n$ bridges.%
\footnote{See Ref.~\cite{DGG95} for an introduction to the subject.}
Neither the river nor the road intersects with itself.
These principle meander numbers, $M_n$, obviously describe the number of
different foldings of a closed strip of $2n$ stamps or of a closed
polymer chain.

One can consider also a generalized problem of the multi-component meander
numbers $M_n^{(k)}$ which are associated with $k$ closed loops of the road
so that $M_n\equiv M_n^{(1)}$. The results of a computer enumeration
of the meander numbers are presented in Refs.~\cite{LZ93,DGG95} up to $n=12$.

\subsection{Hermitean matrix model for meanders}

Meanders can be described by the following Hermitean matrix model~\cite{KK}
\be
{\cal F}_{N\times N}(c)= \frac 2{N^2}
\int \prod_{a=1}^m dW_a \e^{-\frac N2 \sum_{a=1}^m \tr{W_a^2}}
\ln{\left(\int d\phi \e^{-\frac N2 \tr{\phi^2}
+\frac {cN}2 \sum_{a=1}^m \tr{\left(\phi W_a \phi W_a\right)}}\right)}
\label{Hpartition}
\ee
where the integration goes over the $N\times N$ Hermitean matrices
$W_a$ ($a=1,\ldots,m$) and $\phi$. The logarithm in \eq{Hpartition}
leaves only one closed loop of the field $\phi$. The coupling constant
$c$ is associated with the (quartic) interaction between $W_a$ and $\phi$.

Expanding the generating function~\rf{Hpartition} in $c$ and identifying
the diagrams with the ones for the meanders, one relates the large-$N$ limit
of ${\cal F}_{N\times N}(c)$ with the following sum over the meander numbers
\bea
\lim_{N\ra\infty} {\cal F}_{N\times N}(c) =
 \sum_{n=1}^\infty \frac {c^{2n}}{2n} \sum _{k=1}^n M_n^{(k)} m^k \,.
\label{Zm}
\eea
The $N\ra\infty$ limit is needed to keep only planar diagrams as in the meander
problem.

The RHS of \eq{Hpartition} can be expressed entirely via the Gaussian
averages of $W$'s. This leads to the following representation
of the meander numbers:
\be
\sum _{k=1}^n M_n^{(k)} m^k = \sum_{a_1,a_2,\cdots, a_{2n-1}, a_{2n}=1}^m
\LA \frac 1N \tr{W_{a_1} W_{a_2} \cdots W_{a_{2n-1}} W_{a_{2n}}}
\RA^2_{\hbox{\footnotesize{Gauss}}} \,,
\label{words}
\ee
where the average over $W$'s is calculated with the Gaussian weight ---
the same as in~\rf{Hpartition}.
This formula can be proven by calculating the Gaussian integral over $\phi$
in \eq{Hpartition}, expanding the result in $c$ and comparing with
the RHS of \eq{Zm}. The factorization at large $N$ is also used.

The principle meander numbers $M_n$ are given by~\eq{words}
as the linear-in-$m$-terms, \ie as linear terms of the expansion in $m$.
This looks like the replica trick
which suppresses higher loops of the field $W$.

The ordered but cyclic-symmetric sequence of indices
$a_1,a_2,\ldots, a_{2n-1}, a_{2n}$ is often called a {\em word\/}
constructed
of $m$ letters. The average on the RHS of \eq{words} is the meaning
of a word. Thus, the meander problem is equivalent to summing
the squares of all the words with the Gaussian meaning.

The Gaussian averages on the RHS of \eq{Zm} can be represented, making the
Wick pairing, via the Kronecker deltas:
\be
\LA \frac 1N \tr{W_{a_1} W_{a_2} \cdots W_{a_{2n-1}} W_{a_{2n}}}
\RA_{\hbox{\footnotesize{Gauss}}}=\delta_{a_1a_2}
\delta_{a_3a_4} \cdots \delta_{a_{2n-1}a_{2n}} + ~\hbox{planar permutations}
\,.
\label{deltas}
\ee
The ``planar permutations'' means here that one should sum up over all
the permutations of the indices $a_i$'s which are consistent with the
planarity. This is standard for the large-$N$ limit.

To calculate the meander numbers, one should sum up the square of
the RHS of \eq{deltas} over $a_i$'s as is prescribed by \eq{words}.
This is a convenient practical way of calculating the meander
numbers.

Since for $m=1$
\be
\LA \frac 1N \tr{W^{2n}}\RA_{\hbox{\footnotesize{Gauss}}}=
\frac{(2n)!}{(n+1)!n!}\equiv C_n \,,
\label{Catalan}
\ee
which is known as the Catalan number of the order $n$, one gets from
\eq{words}
\be
\sum _{k=1}^n M_n^{(k)} = C_n^2 \,.
\label{1st}
\ee
This is nothing but the first sum rule of Ref.~\cite{DGG95}.

\subsection{Complex matrix model for meanders}

The combination of deltas on the RHS of \eq{deltas} can be
alternatively represented as the Gaussian average over the complex
matrices:
\bea
\lefteqn{\LA \frac 1N \tr{W_{a_1} W^\dagger_{a_2} \cdots
W_{a_{2n-1}} W^\dagger_{a_{2n}} }
\RA_{\hbox{\footnotesize{Gauss}}} } \non &= &
\int \prod_{a=1}^m dW^\dagger_a dW_a
\e^{- N \sum_{a=1}^m \tr{W^\dagger_a W_a }}
\frac 1N \tr{W_{a_1} W^\dagger_{a_2} \cdots W_{a_{2n-1}} W^\dagger_{a_{2n}}}
\,.
\label{complex}
\eea

The generating function, associated with the representation of the meanders
via the complex matrices, reads
\bea
\lefteqn{{\cal F}(c)= \frac 1{N^2}
\LA \ln{\left(\int d\phi_1 d\phi_2 \e^{-S}\right)}
\RA_{\hbox{\footnotesize{Gauss}}} }\non &\equiv& \frac 1{N^2}
\int \prod_{a=1}^m dW^\dagger_a dW_a \e^{- N \sum_{a=1}^m
\tr{W^\dagger_a W_a}}
\ln{\left(\int d\phi_1 d\phi_2 \e^{-S}\right)}
\label{Cpartition}
\eea
with
\be
S=\frac N2 \tr{\phi^2_1}+\frac N2 \tr{\phi^2_2}
-cN \sum_{a=1}^m \tr{\left(\phi_1 W^\dagger_a \phi_2 W_a\right)} \,.
\label{Caction}
\ee
Here $\phi_1$ and $\phi_2$ are Hermitean while $W_a$ ($a=1,\ldots,m$)
are general complex matrices.

Quite similarly to \eq{Hpartition} where the Hermitean matrix $\phi$
can be represented in a diagonal form
$\phi=\hbox{diag}\left( \Lambda^{(1)},\ldots,\Lambda^{(N)} \right)$,
the matrices $\phi_1$ and $\phi_2$ in Eqs.~\rf{Cpartition}, \rf{Caction}
can always be made diagonal:
\bea
\phi_1=\Lambda_1 \equiv
\hbox{diag}\left( \Lambda^{(1)}_1,\ldots,\Lambda^{(N)}_1 \right), \non
\phi_2=\Lambda_2 \equiv
\hbox{diag}\left( \Lambda^{(1)}_2,\ldots,\Lambda^{(N)}_2 \right).
\label{diaphi12}
\eea
This can be shown representing $\phi_1$ and $\phi_2$ as
\be
\phi_1=\Omega_1^\dagger \Lambda_1 \Omega_1\,,~~~~
\phi_2=\Omega_2^\dagger \Lambda_2 \Omega_2
\ee
and absorbing the unitary matrices $\Omega_1$ and $\Omega_2$ by
the transformation of $W_a$:
\be
W_a \longrightarrow \Omega_2^\dagger W_a \Omega_1\,,~~~~
W_a^\dagger \longrightarrow \Omega_1^\dagger W_a^\dagger \Omega_2\,.
\label{Wtrans}
\ee
The measure $dW_a^\dagger dW_a$ does not change under the
transformation~\rf{Wtrans} since $W_a$ are general complex matrices.

It is convenient to introduce one more generating function
\be
M(c) = c\LA \frac{\int d\phi_1 d\phi_2 \e^{-S}
\frac 1N \tr{\phi_1 W^\dagger_1 \phi_2 W_1}}
{\int d\phi_1 d\phi_2 \e^{-S}}
\RAG
\label{defM}
\ee
where only one component of $W_a$, say the first one, enters the averaging
expression. Differentiating the generating function~\rf{Cpartition}
with respect to $c$ and noting that all $m$ components of $W_a$ are
on equal footing, we get the relation
\be
c\frac{d{\cal F}(c)}{dc} = m M(c)
\label{relFM}
\ee
between the two generating functions.

In order to show how the complex matrix model recovers
the meander numbers, let us
 replace $\phi^{ij}_1$ or $\phi^{ij}_2$ in the numerator of \eq{defM} by
 $N^{-1}\partial/\partial \phi^{ji}_1$ or
$N^{-1}\partial/\partial \phi^{ji}_2$, respectively, and integrate by parts.
Repeating this procedure iteratively, we get
\be
M(c)= \sum_{n=1}^\infty c^{2n}
\sum _{k=1}^n M_n^{(k)} m^{k-1}  \,,
\label{Cm}
\ee
with
\be
\sum _{k=1}^n M_n^{(k)} m^{k-1} = \sum_{a_2,\cdots, a_{2n-1}, a_{2n}=1}^m
\LA \frac 1N \tr{W_{1} W^\dagger_{a_2} \cdots
W_{a_{2n-1}} W^\dagger_{a_{2n}}}
\RA^2_{\hbox{\footnotesize{Gauss}}} \,,
\label{Cwords}
\ee
where the Gaussian average is defined by \eq{complex}.
Equation~\rf{Cwords} can be alternatively derived calculating
the Gaussian integrals over $\phi_1$ and $\phi_2$ in \eq{defM} by
virtue of
\bea
\int d\phi_1 d\phi_2 \e^{-S} &=&
\int d\phi_2 \e^{-\fr N2 \tr{\phi^2_2}
+\fr 12 c^2N \sum_{a,b=1}^m \tr{\left(\phi_2 W_a W^\dagger_b \phi_2 W_b
W^\dagger_a\right)}} \non
&=& \det{}^{-1/2}\left[ {\rm I}\otimes {\rm I} -c^2
\sum_{a,b=1}^m  W_a W^\dagger_b \otimes
\left(W_b W^\dagger_a \right)^{\rm t }\right] \,.
\label{cdeterminant}
\eea

For $m=1$ the formula
\be
\LA \frac 1N \tr{(W W^\dagger)^{n}}\RA_{\hbox{\footnotesize{Gauss}}}=
 C_n \,,
\label{cCatalan}
\ee
which is analogous to \eq{Catalan}, holds for the complex matrices.
This results again in \eq{1st}.

It is instructive to consider also the case when $W$ is a fermionic Grassmann
valued matrix {\it \'a la}\/ Ref.~\cite{MZ94}%
\footnote{See Ref.~\cite{SS95} for a review}.
We shall denote the fermionic
matrix as $F$ and its conjugate as $\bar{F}$. Then we get~\cite{MZ94}
\be
\LA \frac 1N \tr{\left(F \bar{F}\right)^{n}}
\RA_{\hbox{\footnotesize{Gauss}}}=
\left\{
\begin{array}{cl}
0 & n=2p ~\hbox{(even)} \\
C_p & n=2p+1 ~\hbox{(odd)}
\end{array}
\right. \,.
\label{fCatalan}
\ee
Since each loop of the fermionic field is accompanied by a factor of $(-1)$,
we arrive at the sum rule
\be
\sum _{k=1}^n  (-)^{k-1} M_n^{(k)}=
\left\{
\begin{array}{cl}
0 & n=2p ~\hbox{(even)} \\
C^2_p & n=2p+1 ~\hbox{(odd)}
\end{array}
\right. \,.
\label{2nd}
\ee
This is nothing but the second sum rule of Ref.~\cite{DGG95}.

Note that the trace of the square of a fermionic matrix vanishes
because of the anticommutation relation imposed on the components.
This is why we did not consider Hermitean fermionic matrices and
used first a representation
of meanders in terms of complex matrices to discuss fermionic representation
of meanders. Fermionic matrix models are a natural representation of
the notion of the signature of arch configurations of Ref.~\cite{DGG95}.

\subsection{General matrix model for meanders}

The consideration of the previous Subsection suggests the
following representation of the meander numbers via a general complex matrix
model which includes both bosonic and fermionic matrices.

Let us consider a general $W_a$ which involves both bosonic (complex) and
fermionic (Grassmann) components:
\bea
W_a= \left( B_1,B_2,\ldots,B_{m_b},F_1,F_2,\ldots,F_{m_f}\right), \non
\bar{W}_a 
= \left( B^\dagger_1,B^\dagger_2,\ldots,B^\dagger_{m_b},
\bar{F}_1,\bar{F}_2,\ldots,\bar{F}_{m_f}\right).
\label{GcontentW}
\eea
Here $m_b$ and $m_f$ ($m=m_b+m_f$)
are the numbers of the bosonic and fermionic components, respectively.
Let us define the generating function ${\cal F}(c)$ by the formulas
\bea
\lefteqn{{\cal F}(c)= \frac 1{N^2}
\LA \ln{\left(\int d\phi_1 d\phi_2 \e^{-S}\right)}
\RA_{\hbox{\footnotesize{Gauss}}} }\non &\equiv& \frac 1{N^2}
\int \prod_{a=1}^m d\bar{W}_a dW_a \e^{- N \sum_{a=1}^m
\tr{\bar{W}_a W_a}}
\ln{\left(\int d\phi_1 d\phi_2 \e^{-S}\right)}
\label{Gpartition}
\eea
and
\be
S=\frac N2 \tr{\phi^2_1}+\frac N2 \tr{\phi^2_2}
-cN \sum_{a=1}^m \tr{\left(\phi_1 \bar{W}_a \phi_2 W_a\right)} \,,
\label{Gaction}
\ee
which generalize Eqs.~\rf{Cpartition} and \rf{Caction}.
Then, the generating function~\rf{Gpartition} is related to the meander
numbers by
\bea
\lim_{N\ra\infty} {\cal F}_{N\times N}(c) = \sum_{n=1}^\infty
\frac {c^{2n}}{2n} \sum _{k=1}^n M_n^{(k)} (m_b-m_f)^k \,.
\label{mf}
\eea
Here $m_f$ emerges with the minus sign since fermion loops are
always accompanied with the minus sign.

Equation~\rf{Cwords} is extended to the given general case of both bosonic
and fermionic matrices as
\bea
\lefteqn{\sum _{k=1}^n M_n^{(k)} (m_b-m_f)^{k-1}
  =  \sum_{a_2,\cdots, a_{2n-1}, a_{2n}=1}^m } \non
&\times& \LA \frac 1N \tr{W_{1} \bar{W}_{a_2} \cdots W_{a_{2n-1}}
\bar{W}_{a_{2n}}} \RAG
 \LA \frac 1N \tr{\bar{W}_{a_{2n}} W_{a_{2n-1}} \cdots \bar{W}_{a_2}
W_{1}} \RAG\,.~~~~
\label{Gwords}
\eea
Analogously, Eqs.~\rf{defM}, \rf{relFM} and \rf{Cm} are generalized as
\be
M(c) = c\LA \frac{\int d\phi_1 d\phi_2 \e^{-S}
\frac 1N \tr{\phi_1 \bar{W}_1 \phi_2 W_1}}
{\int d\phi_1 d\phi_2 \e^{-S}}\RAG \,,
\label{GdefM}
\ee
\be
c\frac{d{\cal F}(c)}{dc} = (m_b-m_f) M(c)\,,
\label{relFMF}
\ee
where
\be
M(c)= \sum_{n=1}^\infty c^{2n} \sum _{k=1}^n M_n^{(k)} (m_b-m_f)^{k-1}\,.
\label{Gm}
\ee
Equations~\rf{defM}, \rf{relFM} and \rf{Cm} are obviously reproduced when
$m_f=0$.

In order to prove Eqs.~\rf{Gwords}, \rf{Gm}, let us first note that
\eq{cdeterminant} is extended to the general case of both bosonic and
fermionic components as
\bea
\int d\phi_1 d\phi_2 \e^{-S} &=&
\int d\phi_2 \e^{-\fr N2 \tr{\phi^2_2}
+\fr 12 c^2N \sum_{a,b=1}^m \tr{\left(\bar{W}_a \phi_2 W_a \bar{W}_b
\phi_2 W_b\right)}} \non
&=&
\int d\phi_2 \e^{-\fr N2 \tr{\phi^2_2}
+\fr 12 c^2N \sum_{a,b=1}^m \sigma (a)
\tr{\left(\phi_2 W_a \bar{W}_b \phi_2
W_b \bar{W}_a\right)}} \non &=&\det{}^{-1/2} \left[ {\rm I}\otimes
{\rm I} -c^2 \sum_{a,b=1}^m \sigma (a) W_a \bar{W}_b \otimes \left(W_b
\bar{W}_a \right)^{\rm t }\right] \,,
\label{Gdeterminant}
\eea
where
\be
\sigma(a)=\left\{
\begin{array}{rl}
1 &~~~~\hbox{for}~~B  \\ -1 &~~~~\hbox{for}~~F
\end{array}
\right.
\label{defsigma}
\ee
is the signature factor of the component $W_a$.

Expanding the determinant~\rf{Gdeterminant} on the RHS of
\eq{Gpartition} in $c^2$,
we get the following representation
\bea
\lefteqn{ \lim_{N\ra\infty} {\cal F}_{N\times N}(c)
 =\sum_{n=1}^\infty \frac {c^{2n}}{2n}\sum_{a_1,\cdots, a_{2n-1},
a_{2n}=1}^m \sigma (a_1)\sigma (a_3)\cdots \sigma (a_{2n-1})}
\non
&\times & \LA \frac 1N \tr{W_{a_1} \bar{W}_{a_2} \cdots W_{a_{2n-1}}
\bar{W}_{a_{2n}}} \RAG
 \LA \frac 1N \tr{W_{a_{2n}} \bar{W}_{a_{2n-1}} \cdots W_{a_2}
\bar{W}_{a_1}} \RAG
\nonumber \\ & & =
\sum_{n=1}^\infty \frac {c^{2n}}{2n}
\sum_{a_1,\cdots, a_{2n-1},
a_{2n}=1}^m  \nonumber \\
& \times &\LA \frac 1N \tr{W_{a_1} \bar{W}_{a_2} \cdots W_{a_{2n-1}}
\bar{W}_{a_{2n}}} \RAG
 \LA \frac 1N \tr{\bar{W}_{a_{2n}} W_{a_{2n-1}} \cdots \bar{W}_{a_2}
W_{a_1}} \RAG
\eea
where the signs of the fermionic components are transformed at the
last step using
\bea
\lefteqn{\sigma(a_1) \sigma(a_3) \cdots \sigma(a_{2n-1})
\LA \frac 1N \tr{W_{a_1} \bar{W}_{a_2} \cdots
W_{a_{2n-1}} \bar{W}_{a_{2n}}}
\RAG } \non &=&
\LA \frac 1N \tr{\bar{W}_{a_1} W_{a_3} \cdots
\bar{W}_{a_{2n-1}} W_{a_{2n}}} \RAG \,.
\label{signatures}
\eea
Therefore, the order of matrices in \eq{Gwords} is chosen in the way
to absorb the signature factors for the fermionic components.

\subsection{Supersymmetric matrix model for principle meander}

Having the representation~\rf{Gwords} of meanders via general
complex matrices (either bosonic or fermionic), we can utilize the
idea of supersymmetry to kill the loops of the $W$-field instead
of the replica trick. Let us consider the two-component $W_a$
whose first component is bosonic while the second one is a fermionic matrix:
\bea
W_a= \left( B,F\right)\,,~~~~ \bar{W}_a \equiv
W^\dagger_a= \left( B^\dagger,\bar{F}\right)\,,
\label{contentW}
\eea
that is given by \eq{GcontentW} with $m_b=m_f=1$.

The generating function~\rf{Gpartition} equals zero for the
supersymmetric model since all the loops of the $B$ and $F$ fields
are mutually cancelled. One should use alternatively the generating
function~\rf{GdefM} which can be represented for the supersymmetric
matrix model as
\be
 M(c) =
\LA \frac 1N \tr{B B^\dagger}
\ln{\left(\int d\phi_1 d\phi_2 \e^{-S}\right)}
\RA_{\hbox{\footnotesize{Gauss}}} \,.
\label{correlator}
\ee
Here $S$ is explicitly given by
\be
S=\frac N2 \tr{\phi^2_1}+\frac N2 \tr{\phi^2_2}
-cN \tr{\left(\phi_1 B^\dagger \phi_2 B\right)}
-cN \tr{\left(\phi_1 \bar F \phi_2 F\right)}
\label{Saction}
\ee
as is prescribed by \eq{Gaction} with $W_a$ substituted
according to \eq{contentW}.
The equivalence of Eqs.~\rf{GdefM}
and \rf{correlator} in the supersymmetric case can be proven
replacing $B$ in the integrand on the RHS of \eq{correlator}
by $N^{-1}\partial/\partial B^\dagger$, integrating by parts,
and recalling that ${\cal F}(c)=0$.

All the multi-component meanders in Eqs.~\rf{Gwords}, \rf{Gm}
vanish in the supersymmetric case
and we get the following representation for the principle meander
\be
M(c)= \sum_{n=1}^\infty c^{2n}  M_n
\label{Sm}
\ee
with
\bea
\lefteqn{M_n = \sum_{a_2,\cdots, a_{2n-1}, a_{2n}=1}^2 } \non & \times &
\LA \frac 1N \tr{B \bar{W}_{a_2} \cdots
W_{a_{2n-1}} \bar{W}_{a_{2n}}}
\RA_{\hbox{\footnotesize{Gauss}}}
\LA \frac 1N \tr{\bar{W}_{a_{2n}}
W_{a_{2n-1}} \cdots  \bar{W}_{a_{2}} B}\RAG
 \,,
\label{Swords}
\eea
where we kept trace of the order of matrices how it appears
from \eq{Gwords}. The signs, which are essential for the fermionic components,
have been transformed using the formula~\rf{signatures} with $\sigma(a)$
being the signature factor of the component $W_a$ defined by \eq{defsigma}.

Equation~\rf{Swords} is a nice representation of
the principle meander numbers 
which looks more natural than the one based on the
replica trick. A hope is that it will be simpler to solve the $m=2$
supersymmetric model than a pure bosonic one at arbitrary $m$.
How the representation~\rf{Swords} reproduces the principle meander
numbers is illustrated by the Table~\ref{t:p.m.} up to $n=4$.
\begin{table}[t]
\centerline{
\begin{tabular}{|c|l|l|l|l|} \hline
$n$ & Structure & Combinatorics & Value & Contribution to $M_n$ \\ \hline
1& $\begin{array}{l} \LA B^2 \RA
   \end{array}$ & $\left. \begin{array}{l} 1
                   \end{array}\right. $ &
                             $\begin{array}{l} 1  \end{array}$ &
                                  $\left. \begin{array}{l} 1
                   \end{array}\right. $\\ \hline
2& $\begin{array}{l} \LA B^4 \RA \\ \LA B^2 F^2 \RA
   \end{array}$ & $\left. \begin{array}{l} 1\\ 2
                   \end{array}\right\} 3$ &
                             $\begin{array}{l} 2 \\ 1 \end{array}$ &
                                  $\left. \begin{array}{l} 4\\ -2
                   \end{array}\right\} 2 $\\ \hline
3& $\begin{array}{l} \LA B^6 \RA \\ \LA B^4 F^2 \RA \\ \LA B^2 F B^2 F\RA \\
\LA B^2 F^4 \RA \\ \LA B F^2 B F^2\RA \\
  \end{array}$ & $\left. \begin{array}{l} 1\\ 4 \\2 \\ 2\\1
                   \end{array}\right\} 10$ &
         $\begin{array}{l} 5 \\ 2 \\1 \\0\\1\end{array}$ &
              $\left. \begin{array}{l} 25\\ -16 \\ -2 \\ 0 \\1
                   \end{array}\right\} 8 $\\ \hline
4& $\begin{array}{l} \LA B^8 \RA \\ \LA B^6 F^2 \RA \\ \LA B^4 F B^2 F\RA \\
\LA B^4 F^4 \RA \\ \LA B^2 F^2 B^2 F^2\RA \\ \LA B^3 F^2 B F^2\RA \\
\LA B^2 F B^2 F^3\RA \\ \LA B^2 F^6 \RA \\ \LA B F^2 B F^4\RA \\
\LA B F^2 B F B^2 F\RA
 \end{array}$ & $\left. \begin{array}{l} 1\\ 6 \\6 \\ 4\\2 \\4\\4\\2\\2\\4
                   \end{array}\right\} 35$ &
         $\begin{array}{l} 14 \\ 5 \\2 \\0\\1\\2\\0\\1\\0\\1\end{array}$ &
       $\left. \begin{array}{l} 196\\ -150 \\ -24 \\ 0 \\2\\16\\0\\-2\\0\\4
                   \end{array}\right\} 42 $ \\ \hline
\end{tabular} }
\caption[x]   {\hspace{0.2cm}\parbox[t]{13cm}
{\small
   Calculation of the principle meander numbers in the supersymmetric
   matrix model~\rf{Swords} up to $n=4$.  }}
\label{t:p.m.}
\end{table}

Alternatively, one can calculate
\bea
\lefteqn{-M_n = \sum_{a_2,\cdots, a_{2n-1}, a_{2n}=1}^2 } \non & \times &
\LA \frac 1N \tr{F \bar{W}_{a_2} \cdots
W_{a_{2n-1}} \bar{W}_{a_{2n}}}
\RA_{\hbox{\footnotesize{Gauss}}}
\LA \frac 1N \tr{\bar{W}_{a_{2n}}
W_{a_{2n-1}} \cdots  \bar{W}_{a_{2}} F}
\RA_{\hbox{\footnotesize{Gauss}}}
 \,.
\label{SFwords}
\eea
The results are presented in the Table~\ref{t:pmf}.
\begin{table}[tb]
\centerline{
\begin{tabular}{|c|l|l|l|l|} \hline
$n$ & Structure & Combinatorics & Value & Contribution to $-M_n$ \\ \hline
1& $\begin{array}{l} \LA F^2 \RA
   \end{array}$ & $\left. \begin{array}{l} 1
                   \end{array}\right. $ &
                             $\begin{array}{l} 1  \end{array}$ &
                                  $\left. \begin{array}{l} -1
                   \end{array}\right. $\\ \hline
2& $\begin{array}{l} \LA F^4 \RA \\ \LA F^2 B^2 \RA
   \end{array}$ & $\left. \begin{array}{l} 1\\ 2
                   \end{array}\right\} 3$ &
                             $\begin{array}{l} 0 \\ 1 \end{array}$ &
                                  $\left. \begin{array}{l} 0\\ -2
                   \end{array}\right\} -2 $\\ \hline
3& $\begin{array}{l} \LA F^6 \RA \\ \LA F^4 B^2 \RA \\ \LA F^2 B F^2 B\RA \\
\LA F^2 B^4 \RA \\ \LA F B^2 F B^2\RA \\
  \end{array}$ & $\left. \begin{array}{l} 1\\ 4 \\2 \\ 2\\1
                   \end{array}\right\} 10$ &
         $\begin{array}{l} 1 \\ 0 \\1 \\2\\1\end{array}$ &
              $\left. \begin{array}{l} -1\\ 0 \\ 2 \\ -8 \\-1
                   \end{array}\right\} -8 $\\ \hline
4& $\begin{array}{l} \LA F^8 \RA \\ \LA F^6 B^2 \RA \\ \LA F^4 B F^2 B\RA \\
\LA F^4 B^4 \RA \\ \LA F^2 B^2 F^2 B^2\RA \\ \LA F^3 B^2 F B^2\RA \\
\LA F^2 B F^2 B^3\RA \\ \LA F^2 B^6 \RA \\ \LA F B^2 F B^4\RA \\
\LA F B^2 F B F^2 B\RA
 \end{array}$ & $\left. \begin{array}{l} 1\\ 6 \\6 \\ 4\\2 \\4\\4\\2\\2\\4
                   \end{array}\right\} 35$ &
         $\begin{array}{l} 0 \\ 1 \\0 \\0\\1\\0\\2\\5\\2\\1\end{array}$ &
       $\left. \begin{array}{l} 0\\ -6 \\ 0 \\ 0 \\2\\0\\16\\-50\\-8\\4
                   \end{array}\right\} -42 $ \\ \hline
\end{tabular} }
\caption[x]   {\hspace{0.2cm}\parbox[t]{13cm}
{\small
   Same as in the Table~\ref{t:p.m.} but using \eq{SFwords}.  }}
\label{t:pmf}
\end{table}

The analogous results for the pure bosonic case given by \eq{Cwords}
at $m=2$ are presented in the Table~\ref{t:Cm2}.
\begin{table}[tb]
\centerline{
\begin{tabular}{|c|l|l|l|l|} \hline
$n$ & Structure & Combinatorics & Value & Contribution \\ \hline
1& $\begin{array}{l} \LA A^2 \RA
   \end{array}$ & $\left. \begin{array}{l} 1
                   \end{array}\right. $ &
                             $\begin{array}{l} 1  \end{array}$ &
                                  $\left. \begin{array}{l} 1
                   \end{array}\right. $\\ \hline
2& $\begin{array}{l} \LA A^4 \RA \\ \LA A^2 B^2 \RA
   \end{array}$ & $\left. \begin{array}{l} 1\\ 2
                   \end{array}\right\} 3$ &
                             $\begin{array}{l} 2 \\ 1 \end{array}$ &
                                  $\left. \begin{array}{l} 4\\ 2
                   \end{array}\right\} 6 $\\ \hline
3& $\begin{array}{l} \LA A^6 \RA \\ \LA A^4 B^2 \RA \\ \LA A^2 B A^2 B\RA \\
\LA A^2 B^4 \RA \\ \LA A B^2 A B^2\RA \\
  \end{array}$ & $\left. \begin{array}{l} 1\\ 4 \\2 \\ 2\\1
                   \end{array}\right\} 10$ &
         $\begin{array}{l} 5 \\ 2 \\1 \\2\\1\end{array}$ &
              $\left. \begin{array}{l} 25\\ 16 \\ 2 \\ 8 \\1
                   \end{array}\right\} 52 $\\ \hline
4& $\begin{array}{l} \LA A^8 \RA \\ \LA A^6 B^2 \RA \\ \LA A^4 B A^2 B\RA \\
\LA A^4 B^4 \RA \\ \LA A^2 B^2 A^2 B^2\RA \\ \LA A^3 B^2 A B^2\RA \\
\LA A^2 B A^2 B^3\RA \\ \LA A^2 B^6 \RA \\ \LA A B^2 A B^4\RA \\
\LA A B^2 A B A^2 B\RA
 \end{array}$ & $\left. \begin{array}{l} 1\\ 6 \\6 \\ 4\\2 \\4\\4\\2\\2\\4
                   \end{array}\right\} 35$ &
         $\begin{array}{l} 14 \\ 5 \\2 \\4\\3\\2\\2\\5\\2\\1\end{array}$ &
   $\left. \begin{array}{l} 196\\ 150 \\ 24 \\ 64 \\18\\16\\16\\50\\8\\4
                   \end{array}\right\} 546 $ \\ \hline
\end{tabular} }
\caption[x]   {\hspace{0.2cm}\parbox[t]{13cm}
{\small
   Same as in the Tables~\ref{t:p.m.} and \ref{t:pmf}
  but using \eq{Cwords} at $m=2$. The components of
  $W_a$ are denoted as $W_1=A,W_2=B$. }}
\label{t:Cm2}
\end{table}

The total number of nonvanishing terms on the RHS of \eq{Cwords}
for the pure bosonic case,
which we shall denote as $\#_n$, is given by the following generating
function
\be
\sum_{n=0}^\infty \#_n c^{2n} =
\frac{\frac m2 \sqrt{1-4(m-1)c^2}-\frac m2 +1}{1-c^2m^2} \,.
\label{numgraphs}
\ee
This formula is derived in the Appendix~A using
non-commutative free random variables.

For $m=2$ \eq{numgraphs} yields
\be
\sum_{n=0}^\infty \#_n c^{2n} =
\frac 1{\sqrt{1-4c^2}} = \sum_{n=0}^\infty
\frac{(2n)!}{(n!)^2} c^{2n}  \,.
\label{numgraphs2}
\ee
These numbers describe the sum of the combinatorial
numbers in the third column of the Tables~\ref{t:p.m.} -- \ref{t:Cm2}.

\subsection{Relation to the Kazakov--Migdal model}

Equation~\rf{numgraphs} is known from the solution of the Kazakov--Migdal
model with the Gaussian potential~\cite{Gro92}. There is the following
reason for that. Suppose that the matrices $W_a$ are unitary instead of
the general complex ones. Then one has
\be
\LA \frac 1N \tr{U_{a_1} U^\dagger_{a_2} \cdots
U_{a_{2n-1}} U^\dagger_{a_{2n}}}
\RA_{\hbox{\footnotesize{Haar measure}}} =
\left\{
\begin{array}{cl}
1 & \hbox{for closed loops} \\
0 & \hbox{for open loops}
\end{array}
\right. \,.
\label{meaning}
\ee
Here the loops represent the sequences of indices
$\{a_1,a_2,\ldots, a_{2n-1}, a_{2n} \}$. The nonvanishing result is only
when the loop is closed and encloses a surface of the vanishing minimal area,
\ie each link of the loop is passed at least twice.
This is analogous to the so-called local confinement in the
Kazakov--Migdal model.

The generating function~\rf{numgraphs} coincides with the following
correlator in the Kazakov--Migdal model with the Gaussian potential
on an infinite $D$-dimensional lattice:
\bea
\lefteqn{\sum_{n=0}^\infty \#_n c^{2n} =
\LA \frac 1N \tr{\phi^2(0)} \RA  }\non &\equiv & \frac{
\int \prod_x d\phi(x)\prod_{\mu=1}^D dU_\mu(x)
 \e^{-S\left[\phi,U\right]}\frac 1N \tr{\phi^2(0)} }
{\int \prod_x d\phi(x)\prod_{\mu=1}^D  dU_\mu(x)
 \e^{-S\left[\phi,U\right]} }
\label{KMcorrelator}
\eea
with the action
\be
S\left[\phi,U\right] =
N \sum_x \left( \fr 12 \tr{\phi^2_x}
-  c \sum_{\mu=1}^D \tr{\left(\phi(x){U}^\dagger_\mu(x) \phi(x+\hat{\mu})
 U_\mu(x)\right)} \right) \,,
\label{KMaction}
\ee
provided that $2D=m$.
The integration over the unitary matrices $U_\mu(x)$ goes over the Haar
measure.

The solution of the
Kazakov--Migdal model with the Gaussian potential can be completely
reformulated as a combinatorial problem of summing over
all closed loops of a given length with all possible backtrackings
(or foldings) included. Its solution~\cite{Mak92} is given by \eq{numgraphs}.

By virtue of the Eguchi--Kawai reduction~\cite{EK82}%
\footnote{See Ref.~\cite{Das87} for a review.},
the correlator~\rf{KMcorrelator} in
the Kazakov--Migdal model on the infinite lattice is equivalent
to that in the reduced model given by
\bea
\sum_{n=0}^\infty \#_n c^{2n} =  \frac{
\int d\phi_1 d\phi_2 \prod_{a=1}^m dU_a
 \e^{-S\left[\phi,U\right]}\frac 1N \tr{\phi^2_1} }
{\int d\phi_1 d\phi_2 \prod_{a=1}^m  dU_a
 \e^{-S\left[\phi,U\right]} },
\label{KMrcorrelator}
\eea
with $m=2D$ and the reduced action being
\be
S\left[\phi,U\right] =
 \fr N2 \tr{\phi^2_1}  + \fr N2 \tr{\phi^2_2}
-  cN \sum_{a=1}^m \tr{\left(\phi_1{U}^\dagger_a \phi_2
 U_a\right)}  \,.
\label{KMraction}
\ee
We have introduced here the index $a$ running from $1$ to $m=2D$ for the
reduced model to distinguish from the index $\mu$ running from $1$ to $D$ for
the Kazakov--Migdal model on the infinite lattice.
The representation~\rf{KMrcorrelator}, \rf{KMraction} can be 
finally rewritten as
\be
\sum_{n=1}^\infty \#_n c^{2n}
 = c\LA \frac{\int d\phi_1 d\phi_2 \e^{-S\left[\phi,U\right]}
\sum_{a=1}^m \frac 1N  \tr{\phi_1 U^\dagger_a \phi_2 U_a}}
{\int d\phi_1 d\phi_2 \e^{-S\left[\phi,U\right]}}
\RA_{\hbox{\footnotesize{Haar measure}}} \,.
\label{defrM}
\ee

In order to prove the equivalence of Eqs.~\rf{KMrcorrelator} and \rf{defrM},
we calculate the integral over $d\phi_1$ and $d\phi_2$ using
\eq{cdeterminant} with $W_a$ substituted by $U_a$:
\bea
\int d\phi_1 d\phi_2 \e^{-S} =
\det{}^{-1/2}\left[ {\rm I}\otimes {\rm I} -c^2
\sum_{a,b=1}^m  U_a U^\dagger_b \otimes
\left(U_b U^\dagger_a \right)^{\rm t }\right] .
\label{detU}
\eea
The determinants in
the numerator and the denominator on the RHS of \eq{defrM}
obviously cancel before the averaging over $U_a$'s. An analogous
cancellation happens in \eq{KMrcorrelator} as well in spite of the fact
that each of them is averaged over its own $U_a$'s. The point is that
the determinant~\rf{detU} under the sign of averaging over $U$
behaves at large $N$ as a $U$-independent constant.

The representation~\rf{defrM} looks very similar to
the generating function~\rf{defM} of the meander numbers. The difference
is that the average is over the unitary matrices in \eq{defrM} and
over the Gaussian complex matrices in \eq{defM}.

We can interpolate between the two cases by modifying the
weight for averaging over $W$'s along the line
of Ref.~\cite{IMS86}. Let us introduce
\bea
\LA F\left[ W,W^\dagger \right]\RA_\alpha \equiv
\int \prod_{a=1}^m \left(dW_a^\dagger dW_a
\e^{-\fr{\alpha N}2\tr{\left({W}^\dagger_a W_a-1+\fr 1\alpha\right)^2}
+\fr N{2\alpha}} \right)
F\left[ W,W^\dagger \right] \,.
\eea
Then the averaging over the Gaussian complex matrices is
reproduced as $\alpha\ra0$ while the average over the unitary
matrices is recovered as $\alpha\ra\infty$ since
the matrix $W_\a$ is forced to be unitary as $\alpha\ra\infty$.

We see, thus, that the words are the same both for the meander problem
and for the Kazakov--Migdal model. The only difference resides in the meaning
of nonvanishing words --- it equals unity for the unitary matrices.

\newsection{Representation via non-commutative variables}

The set $\u_a$, $\ud_a$ of non-commutative variables obey the Cuntz algebra
\be
\u_a \ud_b = \delta_{ab} \,.
\label{Cuntz}
\ee
It is convenient to consider them, respectively,
as annihilation and creation operators in a Hilbert space with the vacuum
$\left|\Omega \RA$ which satisfies
\bea
\u_a \left|\Omega \RA =0\,,~& &~\LA \Omega \right| \ud_a=0\,, \non
\LA \Omega|\Omega \RA &=&1 \,.
\label{vacuum}
\eea
The completeness condition says that
\be
\sum_{a=1}^m \ud_a \u_a =1 - \left|\Omega \RA \LA \Omega \right| \,.
\label{completeness}
\ee
There are no more relations between the non-commutative variables.

\subsection{Bosonic case}

Let us construct the generating function for words via
the non-commutative sources $\u_a$ as
\bea
G_\l(\u)&= &\LA \ntr \frac{1}{\l-\sum_{a=1}^m \u_a W_a} \RAG \non &= &
\sum_{n=0}^\infty \frac{1}{\l^{2n+1}}
 \sum_{a_1,a_2,\cdots,a_{2n}=1}^m  \u_{a_1}\u_{a_2} \cdots \u_{a_{2n}}
\LA \ntr W_{a_1} W_{a_2} \cdots W_{a_{2n}} \RAG
\label{defG}
\eea
where the average over $a=1,\ldots,m$ matrices $W_a$ is
with the Gaussian weight as before.

Then the generating function~\rf{defM} for the meander numbers
is given by
\be
\LA\Omega |G_\l(\u) G_\l(\u^\dagger) |\Omega \RA = c+c m M(c)
\label{nondefM}
\ee
with $c=1/\l^2$. The contraction of indices on the RHS of \eq{words}
is obviously reproduced using Eqs.~\rf{Cuntz} and \rf{vacuum}.

The generating function $G_\l(\u)$ obeys the Schwinger-Dyson equation
\bea
\l G_\l(\u) -1 = \sum_{a=1}^m G_\l(\u)\u_a G_\l(\u) \u_a \,,
\label{SD}
\eea
which can be derived in a usual way, by shifting $W$.
The cyclic symmetry of the trace implies
\bea
\l G_\l(\u) -1 = \sum_{b=1}^m \u_b \left( \l G_\l(\u)-1\right) \ud_b \,.
\label{cySD}
\eea
Inserting here \eq{SD}, we get
\bea
\sum_{a=1}^m G_\l(\u)\u_a G_\l(\u) \u_a =
\sum_{a=1}^m \u_a G_\l(\u)\u_a G_\l(\u) \,.
\label{cyclic}
\eea
An alternative combinatorial derivation of Eqs.~\rf{nondefM} and \rf{SD} is
presented in the Appendix~B.

We shall use for $G_\l(\u)$ and $G_\l(\ud)$ the short-hand notation
\be
G_\l \equiv G_\l(\u)\,,~~~G^\dagger_\l \equiv G_\l(\ud) \,,
\label{defGd}
\ee
so that \eq{SD} can be rewritten as
\bea
\l G_\l -1 &=& G_\l\u_a G_\l \u_a \,, \non
\l G_\l^\dagger -1 &= &\ud_a G_\l^\dagger\ud_a G_\l^\dagger \,,
\label{leq}
\eea
where the summation over repeated indices is implied here and
below except when it is specially indicated.
Using \eq{Cuntz}, one alternatively rewrites \eq{leq} as
\bea
\left(\l G_\l -1 \right) \ud_a &= &G_\l \u_a G_\l \,, \non
\u_a \left(\l G_\l^\dagger -1 \right)& =& G_\l^\dagger \ud_a G_\l^\dagger \,.
\label{SDa}
\eea

If $u$'s were ordinary commutative variables, the solution to the quadratic
equation~\rf{SD} would be simple
\be
G_\l=\frac{\l-\sqrt{\l^2-4\vec{\,u}^2}}{2\vec{\,u}^2} \,.
\ee
For $m=1$ this is nothing but Wigner's semicircle law and
one reproduces \eq{cCatalan} by expanding in $1/\l$.

A formal solution for the non-commutative variables can be obtained
representing \eq{SD} as
\be
G_\l= \frac{1}{\l-\u_a G_\l \u_a}.
\label{iSD}
\ee
Iterations of this equation, as was found by Cvitanovi\'{c}~\cite{Cvi81},
lead in the continued fraction
\bea
G_\l\left(\u \right)=\frac{1}{\displaystyle \l-\u_{a_1}\frac{1}
{\displaystyle \l-\u_{a_2}\frac{1}{\displaystyle \l-\u_{a_3}
\frac{1}{\displaystyle \vdots}\,\u_{a_3}}\,\u_{a_2}}\,\u_{a_1}}\,.
\label{Cvi}
\eea
Expanding the RHS in $1/\l^2$, one gets
\bea
G_\l\left(\u \right)=\frac{1}{\l}+\frac{\u^2}{\l^3}
+\frac{\u^2\u^2+\u_a\u^2\u_a}{\l^5} \hspace*{6cm}\non +
\frac{\u^2\u^2\u^2+\u_a\u^2\u^2\u_a+\u^2\u_a\u^2\u_a+
\u_a\u^2\u_a\u^2+\u_a\u_b \u^2 \u_b \u_a}{\l^7}
+{\cal O}\left(\frac1{\l^9} \right)
\label{nonexpansion}
\eea
where $\u^2$ stands for $\u_a\u_a$. All possible planar combinations of
$\u$'s appear to next orders (with unit coefficients) while the total
number of terms to the order $\l^{-2n-1}$ equals the Catalan number $C_n$.
Technically, it is more simple to derive the expansion of
$G_\l\left(\u \right)$ in $1/\l$ by direct iterations of \eq{SD}.
The substitution into \eq{nondefM} then recovers the lower meander numbers.

Equation~\rf{SD} is in fact well-known for the Gaussian models.
It is a consequence of the relation between the generating functionals
for all planar graphs $G_\l\left(\u \right)$ and for connected planar
graphs
\bea
{\cal W}\left(\j\right)&= &\LA \ntr \frac{1}{\l-\sum_{a=1}^m \j_a W_a}
\RA_{\rm conn}\non &= &
\sum_{n=0}^\infty \frac{1}{\l^{2n+1}}
 \sum_{a_1,a_2,\cdots,a_{2n}=1}^m  \j_{a_1}\j_{a_2} \cdots \j_{a_{2n}}
\LA \ntr W_{a_1} W_{a_2} \cdots W_{a_{2n}} \RA_{\rm conn}
\eea
which says~\cite{BIPZ78,Cvi81}
\be
G\left(\u \right)= {\cal W}\left(\l\u G\left(\u \right)\right)
\label{GSD}
\ee
while the cyclic symmetry gives
\be
{\cal W}\left(\l\u G\left(\u \right)\right)=
{\cal W}\left(\l G\left(\u \right) \u\right).
\label{Gcyclic}
\ee
There is only one connected graph in the Gaussian case so that
${\cal W}\left(\j\right)$ is quadratic
\be
{\cal W}\left(\j\right) = \frac1\l +\frac{\j_a\j_a}{\l^3} \,.
\ee
Equations~\rf{GSD} and \rf{Gcyclic} now recover Eqs.~\rf{SD} and
\rf{cyclic}, respectively.

It would be interesting to apply the theory of non-commutative free
random variables~\cite{VDN92}, whose application to
matrix models has been discussed recently in Refs.~\cite{Dou95,GG95},
for this problem.

\subsection{General case}

For a general complex matrix model when some components of $W_a$ are
bosonic and some are fermionic, we define the generating functions
\bea
\lefteqn{G_\l(\u)= \LA \frac 1N \tr \frac{\lambda}{\lambda^2 -
\sum_{a,b=1}^m \u_a \u_b W_a \bar{W}_b}\RAG =}
\non & &
\sum_{n=0}^\infty \frac{1}{\l^{2n+1}}
 \sum_{a_1,a_2,\cdots,a_{2n}=1}^m
 \u_{a_1}\u_{a_2} \cdots \u_{a_{2n-1}}\u_{a_{2n}}
\LA \ntr W_{a_1} \bar{W}_{a_2} \cdots {W}_{a_{2n-1}}  \bar{W}_{a_{2n}}\RAG
\label{CdefG}
\eea
and
\bea
\lefteqn{\bar{G}_\l(\u)= \LA \frac 1N \tr \frac{\lambda}{\lambda^2 -
\sum_{a,b=1}^m \u_a \u_b \bar{W}_a {W}_b}\RAG = }
\non & &
\sum_{n=0}^\infty \frac{1}{\l^{2n+1}}
 \sum_{a_1,a_2,\cdots,a_{2n}=1}^m
 \u_{a_1}\u_{a_2} \cdots \u_{a_{2n-1}} \u_{a_{2n}}
\LA \ntr\bar{W}_{a_1} {W}_{a_2} \cdots \bar{W}_{a_{2n-1}} {W}_{a_{2n}}\RAG
\label{barCdefG}
\eea
where the non-commutative sources $\u_a$ are the same as before.

The generating functions $G_\l(\u)$ and $\bar{G}_\l(\u)$ are not
independent due to \eq{signatures}. One gets
\be
\bar{G}_\l(\u_a) ={G}_\l(\bar{\u}_a)
\label{GG}
\ee
where
\be
\bar{\u}_a \equiv \sqrt{\sigma(a)} \u_a
\label{baru}
\ee
and
$\sigma(a)$ is defined by \eq{defsigma}. In other words the components
of $\u_a$ which are associated with fermionic components of
$W_a$ should be multiplied by $i$ on the RHS of \eq{GG} while the
ones for the bosonic components remain unchanged.

The Schwinger-Dyson equations for $G_\l(\u)$ and $\bar{G}_\l(\u)$ can
be derived in a usual way from the recurrence relations
\bea
\lefteqn{\LA \ntr W_{a_1} \bar{W}_{a_2} \cdots {W}_{a_{2n-1}}
\bar{W}_{a_{2n}}\RAG }
\non &=& \sum_{k=0}^{n-1} \delta_{a_{2n}a_{2k+1}}
\LA \ntr W_{a_1} \bar{W}_{a_2} \cdots   \bar{W}_{a_{2k}}\RAG
\LA \ntr \bar{W}_{a_{2k+2}} \cdots {W}_{a_{2n-1}} \RAG
\label{recurrence1}
\eea
and
\bea
\lefteqn{
\LA \ntr W_{a_1} \bar{W}_{a_2} \cdots {W}_{a_{2n-1}}  \bar{W}_{a_{2n}}\RAG }
\non &= &\sum_{k=1}^{n} \delta_{a_{1}a_{2k}}
\LA \ntr \bar{W}_{a_2} \cdots {W}_{a_{2k-1}} \RAG
\LA \ntr {W}_{a_{2k+1}} \cdots \bar{W}_{a_{2n}} \RAG \,.
\label{recurrence2}
\eea
They can be obtained in a usual way
by shifting $\bar{W}_{2n}$ and $W_1$, respectively.

Equations~\rf{recurrence1} and \rf{recurrence2}
and the cyclic symmetry of the trace result in the equations
\bea
\l G_\l(\u) -1 = \sum_{a=1}^m \u_a  \bar{G}_\l(\u)\u_a {G}_\l(\u)
= \sum_{a=1}^m G_\l(\u)\u_a \bar{G}_\l(\u) \u_a \,,
\label{SSD}
\eea
and
\bea
\l \bar{G}_\l(\u) -1 = \sum_{a=1}^m \sigma(a)
\bar{G}_\l(\u)\u_a G_\l(\u) \u_a =\sum_{a=1}^m \sigma(a)
\u_a {G}_\l(\u)\u_a \bar{G}_\l(\u)\,.
\label{barSSD}
\eea
These two equations are not independent due to the relation~\rf{GG}.
One can be obtained from another by the substitution
\be
\u_a\ra \bar{\u}_a  \,,
\ee
where $\bar{\u}_a$ is given by \eq{baru}. Their combinatorial
interpretation is presented in the Appendix~B.

Introducing in addition to \eq{defGd} the short-hand notations
\be
\bar{G}_\l \equiv \bar{G}_\l(\u)\,,~~~
\bar{G}^\dagger_\l \equiv \bar{G}^\dagger_\l(\ud) \,,
\label{CdefGd}
\ee
one derives from Eqs.~\rf{SSD}, \rf{barSSD} the analog of \eq{SDa} as
\bea
\left(\l G_\l -1 \right) \ud_a &= &G_\l \u_a \bar{G}_\l \,, \non
\u_a \left(\l {G}_\l^\dagger -1 \right)& =& \bar{G}_\l^\dagger
 \ud_a G_\l^\dagger
\label{SSDa}
\eea
and
\bea
\left(\l \bar{G}_\l -1 \right) \ud_a &= &
\sigma(a) \bar{G}_\l \u_a {G}_\l \,, \non
\u_a \left(\l \bar{G}_\l^\dagger -1 \right)& =&
\sigma(a) G_\l^\dagger \ud_a \bar{G}_\l^\dagger \,.
\label{barSSDa}
\eea

The solution to Eqs.~\rf{SSD} and \rf{barSSD} can be easily obtained
for a pure fermionic model and commutative sources.
Summing up Eqs.~\rf{SSD} and \rf{barSSD}, we get
\be
G_\l=\frac2\lambda -\bar{G}_\l
\ee
and
\be
\l \bar{G}_\l -1 = \vec{\,u}^2 \bar{G}^2_\l -\frac2\lambda \bar{G}_\l
\label{EEE}
\ee
with the solution
\be
\bar{G}_\l = \frac 1\l + \frac \l{2\vec{\,u}^2}
-\frac1{2\l\vec{\,u}^2}\sqrt{\l^4+4\vec{\,u}^4}
\label{11}
\ee
and
\be
{G}_\l = \frac 1\l -\frac \l{2\vec{\,u}^2}
+\frac1{2\l\vec{\,u}^2}\sqrt{\l^4+4\vec{\,u}^4} \,.
\label{22}
\ee
Equations~\rf{EEE} and \rf{11} for $m=1$ recover the ones of Ref.~\cite{MZ94}
while the expansion of \eq{22} in $1/\l$ gives \eq{fCatalan}.

In order to find a formal solution to Eqs.~\rf{SSD} and \rf{barSSD}
in the form of a continued fraction, let us rewrite them as
\be
G_\l= \frac{1}{\l-\u_a \bar{G}_\l \u_a}
\label{iSSD}
\ee
and
\be
\bar{G}_\l= \frac{1}{\l-\bar{\u}_a G_\l \bar{\u}_a}
\label{ibarSSD}
\ee
where $\bar{\u}_a$ is defined by \eq{baru}.
Iterations of this equations lead to the following analog of~\rf{Cvi}
\be
G_\l\left(\u \right)=\frac{1}{\displaystyle \l-\u_{a_1}\frac{1}
{\displaystyle \l-\bar{\u}_{a_2}
\frac{1}{\displaystyle \l-\u_{a_3}\frac{1}{\displaystyle \l-\bar{\u}_{a_4}
\frac{1}{\vdots}\, \bar{\u}_{a_4}}\,\u_{a_3}}\,\bar{\u}_{a_2}}\,\u_{a_1}}
\label{SCvi}
\ee
and
\be
\bar{G}_\l\left(\u \right)=\frac{1}{\displaystyle \l-\bar{\u}_{a_1}\frac{1}
{\displaystyle \l-\u_{a_2}\frac{1}{\displaystyle \l-\bar{\u}_{a_3}
\frac{1}{\displaystyle \l-\u_{a_4}\frac{1}{\vdots}
\,\u_{a_4}}\,\bar{\u}_{a_3}}\, \u_{a_2}}\, \bar{\u}_{a_1}} \,.
\label{barSCvi}
\ee
Here $\u$ and $\bar{\u}$ interchange in the consequent lines of the
continued fractions.

The quantities on the LHS's of Eqs.~\rf{SSDa} and \rf{barSSDa} read explicitly
\be
\u_a \left(\l {G}_\l^\dagger -1 \right)=
\sum_{n=1
}^\infty \frac{1}{\l^{2n}}
 \sum_{a_2,\cdots,a_{2n}=1}^m  \u^\dagger_{a_2} \cdots
\u^\dagger_{a_{2n-1}} \u^\dagger_{a_{2n}}
\LA \ntr W_{a} \bar{W}_{a_2} \cdots {W}_{a_{2n-1}}  \bar{W}_{a_{2n}}\RAG
\label{openindex}
\ee
and
\be
\left(\l \bar{G}_\l -1 \right) \ud_a =
\sum_{n=1}^\infty \frac{1}{\l^{2n}}
 \sum_{a_1,a_2,\cdots,a_{2n-1}=1}^m \!\!\!\!\!\!\u_{a_1} \u_{a_2} \cdots
\u_{a_{2n-1}}
\LA \ntr\bar{W}_{a_1} W_{a_2} \cdots \bar{W}_{a_{2n-1}} W_a \RAG .
\label{baropenindex}
\ee

The generating function~\rf{GdefM} is determined by
\be
\LA\Omega |G_\l(\u) \bar{G}_\l(\u^\dagger) |\Omega \RA =
\LA\Omega |\bar{G}_\l(\u) G_\l(\u^\dagger) |\Omega \RA = c+c (m_b-m_f) M(c)
\label{SnondefM}
\ee
where $m_b$ and $m_f$ are the numbers of bosonic and fermionic components
of $W_a$, respectively, and $\l^2=1/c$. This formula follows from
the fact that the LHS reproduces the contraction of indices as in
\eq{Gwords} using Eqs.~\rf{Cuntz} and \rf{vacuum}.

\subsection{Supersymmetric case}

For the supersymmetric case $m_b=m_f=1$ and \eq{SnondefM} does not
determine the meander numbers. One should use, instead, \eq{Swords}
or \eq{SFwords}, where there is no summation over one of the indices,
to get the principle meander numbers.

Let us denote the components of $\u_a$ as
\be
\u_a=\left( u,v\right).
\ee
Equations~\rf{openindex} and \rf{baropenindex} then result in
\be
M(\frac1{\l^2}) = \l^2 \LA\Omega | \bar{G}_\l u^\dagger u {G}_\l^\dagger
 |\Omega \RA = -\l^2 \LA\Omega | \bar{G}_\l v^\dagger v {G}_\l^\dagger
 |\Omega \RA
\label{barsmea}
\ee
or alternatively
\be
M(\frac1{\l^2}) = \l^2 \LA\Omega | {G}_\l u^\dagger u \bar{G}_\l^\dagger
 |\Omega \RA = -\l^2 \LA\Omega | {G}_\l v^\dagger v \bar{G}_\l^\dagger
 |\Omega \RA .
\label{smea}
\ee

The equality sign between the two expressions
on the RHS of \eq{barsmea} or \eq{smea} is due to the supersymmetry.
This can be shown using the completeness condition~\rf{completeness}
which takes in the supersymmetric case the form
\be
u^\dagger u + v^\dagger v=1 - \left|\Omega \RA \LA \Omega \right| \,.
\label{scompleteness}
\ee
Inserting this, say, between $G_\l$ and $\bar{G}_\l^\dagger$, we get
\be
\l^2 {G}_\l u^\dagger u \bar{G}_\l^\dagger
 = -\l^2 {G}_\l v^\dagger v \bar{G}_\l^\dagger +
\l^2 G_\l \bar{G}_\l^\dagger -\left|\Omega \RA \LA \Omega \right|\,.
\ee
Taking the vacuum expectation value of this formula and remembering
that
\be
\l^2 \LA \Omega \right|  G_\l \bar{G}_\l^\dagger \left|\Omega \RA = 1
\label{=1}
\ee
due to the supersymmetry, we prove the statement.

Using Eqs.~\rf{SSDa} and \rf{barSSDa}, Eqs.~\rf{smea} and \rf{barsmea}
 can be rewritten as
\be
M(\frac1{\l^2}) =\LA\Omega | G_\l u\bar{G}_\l
G_\l^\dagger u^\dagger\bar{G}_\l^\dagger |\Omega \RA
=\LA\Omega | G_\l v\bar{G}_\l
G_\l^\dagger v^\dagger\bar{G}_\l^\dagger |\Omega \RA
\label{mea}
\ee
and
\be
M(\frac1{\l^2}) =\LA\Omega | \bar{G}_\l u{G}_\l
\bar{G}_\l^\dagger u^\dagger {G}_\l^\dagger |\Omega \RA
=\LA\Omega | \bar{G}_\l v {G}_\l
\bar{G}_\l^\dagger v^\dagger {G}_\l^\dagger |\Omega \RA .
\label{barmea}
\ee
This representation of the generating function for the
principle meander numbers is convenient for an iterative procedure.

To perform calculations order by order in $c=1/\l^2$, it looks reasonable to
try to use the fact that there are only two non-commutative variables
and to expand in $v$. This is a standard trick of dealing with two
non-commutative variables in the Hermitean two-matrix
model~\cite{DL95,COT95}.

We introduce, therefore, the expansions
\bea
G_\l&=&\sum_{n=0}^\infty G_\l^{(n)} \,, \non
\bar{G}_\l&=&\sum_{n=0}^\infty (-1)^n G_\l^{(n)} \,,
\label{expansion}
\eea
where $(-1)^n$ in the expansion of $\bar{G}_\l$ is due to Eqs.~\rf{GG}
\rf{baru}. We have
\be
G_\l^{(0)}=\frac{\l-\sqrt{\l^2-4u^2}}{2u^2}
\label{Wieg}
\ee
while $G_\l^{(n)}$ involves $v$ exactly $2n$ times. As we shall see below,
\bea
G_\l^{(n)} &\sim& \frac{1}{\l^{2n+1}}~~~~~~\hbox{for $n$ odd}\,,\non
G_\l^{(n)}& \sim &\frac{1}{\l^{2n+3}}~~~~~~\hbox{for $n\geq2$ even}\,.
\label{lexpansion}
\eea
Therefore, a contribution to $M_n$ arises in \eq{smea} at most
from $G_\l^{(n)}$ for odd $n$ and from $G_\l^{(n-1)}$ for even $n$.

The functions $G_\l^{(n)}$ can be found recursively from \eq{iSSD}
which takes the form
\be
\sum_{n=0}^\infty G_\l^{(n)}=G_\l^{(0)}
\frac{1}{1-\sum_{n=1}^\infty (-1)^n uG_\l^{(n)} uG_\l^{(0)}   -
\sum_{n=0}^\infty (-1)^n vG_\l^{(n)} v G_\l^{(0)} } \,.
\label{forrec}
\ee
At each step one has to solve the equation
\be
G_\l^{(n)}=(-1)^n G_\l^{(0)}u G_\l^{(n)} uG_\l^{(0)} + A_n
\ee
with some $A_n$ whose solution is given by
\bea
G_\l^{(2p-1)}&= &\sum_{l=0}^\infty (-1)^l
\left(G_\l^{(0)} u  \right)^l A_{2p-1}\left(G_\l^{(0)} u  \right)^l \,,\non
G_\l^{(2p)}&=& \sum_{l=0}^\infty
\left(G_\l^{(0)} u  \right)^l A_{2p}\left(G_\l^{(0)} u  \right)^l \,.
\label{per}
\eea
where $p\geq1$. Both formulas can be rewritten in a unique way via
the contour integral
\be
G_\l^{(n)}=\oint \frac{d\omega}{2\pi i}
\frac{1}{\left(\omega-(-1)^n G_\l^{(0)}u\right)}
A_n \frac{1}{\left(1-\omega G_\l^{(0)}u \right)} \,.
\label{oper}
\ee
We shall also introduce the obvious short-hand notation for RHS's of
Eqs.~\rf{per} or \rf{oper} as the brackets
\be
G_\l^{(n)}=\lbr A_n \rbr \,.
\label{bra}
\ee

The functions $A_n$ can be found recursively from \eq{forrec}.
Few lower ones read explicitly
\bea
A_1&=& G_\l^{(0)} v G_\l^{(0)} v G_\l^{(0)} \,, \nonumber \\
A_2&=& - G_\l^{(0)} v G_\l^{(1)} v G_\l^{(0)}
+  G_\l^{(1)} \frac{1}{ G_\l^{(0)}}  G_\l^{(1)}\,, \nonumber \\
A_3&= &G_\l^{(0)} v  G_\l^{(2)} v  G_\l^{(0)} +
G_\l^{(2)} \frac{1}{G_\l^{(0)}}  G_\l^{(1)} +
G_\l^{(1)} \frac{1}{G_\l^{(0)}}  G_\l^{(2)}
- G_\l^{(1)} \frac{1}{G_\l^{(0)}}G_\l^{(1)} \frac{1}{G_\l^{(0)}} G_\l^{(1)}
 \,. ~~~
\label{As}
\eea
They obviously satisfy the equation
\be
\l A_n v^\dagger = \sum_{k=0}^{n-1} (-1)^k
G_\l^{(n-1-k)} v G_\l^{(k)}
\ee
and analogously
\be
\l v A^\dagger_n  = \sum_{k=0}^{n-1} (-1)^k
G_\l^{\dagger\,(k)} v^\dagger G_\l^{\dagger\,(n-1-k)}\,,
\ee
which can be obtained substituting the expansion~\rf{expansion}
into \eq{SSDa} (or \eq{barSSDa}) and using the fact that
\be
G_\l^{(n)}v^\dagger = A_n v^\dagger
\label{GvAv}
\ee
due to \eq{per}.

Using \eq{bra}, we get from \eq{As} for $G_\l^{(n)}$'s
\bea
G_\l^{(1)}&=& G_\l^{(0)} \lbr v G_\l^{(0)} v\rbr G_\l^{(0)}  \,, \nonumber \\
G_\l^{(2)}&=& - G_\l^{(0)} \lbr v G_\l^{(0)} \lbr v G_\l^{(0)} v
\rbr G_\l^{(0)}  v \rbr G_\l^{(0)}
+ G_\l^{(0)}  \lbr\lbr v G_\l^{(0)} v \rbr G_\l^{(0)}
 \lbr  v G_\l^{(0)} v \rbr\rbr G_\l^{(0)} \,,
\nonumber \\
G_\l^{(3)}&= &-G_\l^{(0)} \lbr v G_\l^{(0)} \lbr v G_\l^{(0)}\lbr v G_\l^{(0)}
v \rbr G_\l^{(0)} v\rbr G_\l^{(0)}  v \rbr G_\l^{(0)} \non & &
+G_\l^{(0)} \lbr v G_\l^{(0)} \lbr v G_\l^{(0)} v \rbr G_\l^{(0)}
 \lbr v G_\l^{(0)} v \rbr G_\l^{(0)} v \rbr G_\l^{(0)}  \non
& &- G_\l^{(0)} \lbr\lbr  v G_\l^{(0)} v \rbr
 G_\l^{(0)} \lbr v G_\l^{(0)} \lbr v G_\l^{(0)} v
\rbr G_\l^{(0)}  v \rbr \rbr G_\l^{(0)} \non
& & +G_\l^{(0)} \lbr\lbr v G_\l^{(0)} v \rbr
{ G_\l^{(0)}} \lbr v G_\l^{(0)} v  \rbr
{ G_\l^{(0)}} \lbr  v G_\l^{(0)} v \rbr\rbr G_\l^{(0)} \non
& &- G_\l^{(0)} \lbr \lbr  v G_\l^{(0)} \lbr v G_\l^{(0)} v
\rbr G_\l^{(0)}  v  \rbr { G_\l^{(0)}}
\lbr  v G_\l^{(0)} v  \rbr \rbr G_\l^{(0)} \,,
\label{Gs}
\eea
where we have used the fact that
$G_\l^{(0)}$ commutes with a bracket but not
with $v$. There are no problems here with the order of brackets which act like
in TeX: they immediately contract each other when it is possible.
This contraction is associated with the summation~\rf{per}
(or the integration~\rf{oper}).
The general $G_\l^{(n)}$ contains $C_n$ (the Catalan number
given by \eq{Catalan})
terms of this kind with alternating signs. Some rules for
representing a general term can be formulated
which resemble the Wick pairing of bilinear combinations of $v$'s.

It is worth noting that most of the above formulas would look quite similar
for the case of just 2 bosonic fields, \ie of $m=2$, while there will
be no alternative signs for 2 bosons. Due to these minus signs, the
supersymmetric case is somewhat
simpler than the $m=2$ one. Say, the leading-order
terms $v^4$ are cancelled in \eq{As} for $A_2$ and in \eq{Gs} for
$G_\l^{(2)}$. This is a reflection of the general property~\rf{fCatalan} of
Gaussian averages of the fermionic matrices. It is the reason why
the leading order term vanishes in \eq{lexpansion} for even $n$.
In addition some relations are imposed by the supersymmetry.
The simplest one follows from the substitution of the expansion~\rf{expansion}
into \eq{=1} which yields
\be
\l^2 \sum_{n=0}^\infty (-1)^n \LA \Omega \right|
G_\l^{(n)} G_\l^{\dagger \,(n)}\left|\Omega \RA = 1 \,.
\label{sum=1}
\ee
Only the diagonal terms survive here due to the definitions~\rf{Cuntz},
\rf{vacuum}.

In order to calculate $M_n$, we substitute the expansion~\rf{expansion}
into \eq{smea} which gives
\be
M(\frac1{\l^2}) =
\l^2 \sum_{n=1}^\infty (-1)^{(n-1)} \LA \Omega \right|
G_\l^{(n)}v^\dagger v G_\l^{\dagger \,(n)}\left|\Omega \RA \,.
\label{sumsmea}
\ee
We can also rewrite the RHS as
\be
M(\frac1{\l^2}) =
\l^2 \sum_{n=1}^\infty (-1)^{(n-1)} \LA \Omega \right|
A_n v^\dagger v A_n^\dagger \left|\Omega \RA
\label{Asumsmea}
\ee
due to \eq{GvAv}.

It is easy to calculate the term with $n=1$ in \eq{sumsmea}
(or \eq{Asumsmea}). Using \eq{Gs}, we get
\be
G_\l^{(1)} v^\dagger = \frac 1\l G_\l^{(0)} v G_\l^{(0)}\,,~~~~
v G_\l^{\dagger\,(1)} = \frac 1\l G_\l^{\dagger \,(0)}
v^\dagger  G_\l^{\dagger \,(0)} \,,
\ee
so that Eqs.~\rf{Cuntz}, \rf{vacuum} and \rf{Wieg} result in
\be
M(\frac1{\l^2}) = c^2 \left(\sum_{k=0}^\infty
c^{2k} C^2_k \right)^2 +
\l^2 \sum_{n=2}^\infty (-1)^{(n-1)} \LA \Omega \right|
G_\l^{(n)}v^\dagger v G_\l^{\dagger \,(n)}\left|\Omega \RA \,.
\label{0sumsmea}
\ee
The first term on the RHS recovers $M_1=1$ and $M_2=2$ while the contribution
of the next terms are controlled by \eq{lexpansion}.

Let us demonstrate of how this iterative procedure works by the
explicit calculation
up to $n=3$ when at most $G_\l^{(3)}$ is essential. We get from \eq{Gs}
\bea
G_\l^{(2)} v^\dagger = \frac{-v u v^2 u + u v^2 u v}{\l^7}
+{\cal O}\left( \frac{1}{\l^9}\right), & &
v G_\l^{\dagger\,(2)} =
\frac{v^\dagger u^\dagger v^{\dagger\,2} u^\dagger -
 u^\dagger v^{\dagger\,2} u^\dagger v^\dagger}{\l^7}
+{\cal O}\left( \frac{1}{\l^9}\right), \nonumber \\
G_\l^{(3)} v^\dagger = -\frac{v^6}{\l^7}
+{\cal O}\left( \frac{1}{\l^9}\right), & &
v G_\l^{\dagger\,(3)} =
-\frac{ v^{\dagger\,6}}{\l^7}
+{\cal O}\left( \frac{1}{\l^9}\right).
\eea
The substitution into \eq{0sumsmea} recovers $M_3=8$.

A most difficult part of the described general iterative procedure is to
calculate \sloppy $\LA \Omega \right|
G_\l^{(n)} v^\dagger v G_\l^{\dagger\,(n)}
 \left|\Omega \RA$ which involves $C^2_n$ terms for $n>1$.
There is no problem to
calculate a contribution from each individual term by using the formula
\bea
\lefteqn{\LA \Omega \right|
f_0(u)v f_1(u)v \cdots v f_n(u) g_n(u^\dagger)v^\dagger \cdots
v^\dagger g_1(u^\dagger) v^\dagger g_0(u^\dagger)\left|\Omega \RA} \non &=&
\LA \Omega \right| f_0(u)g_0(u^\dagger)\left|\Omega \RA
\LA \Omega \right| f_1(u)g_1(u^\dagger)\left|\Omega \RA
\cdots \LA \Omega \right| f_n(u)g_n(u^\dagger)\left|\Omega \RA \,,
\eea
which is a consequence of Eqs.~\rf{Cuntz} and \rf{vacuum}.
When calculating order by order in $c=1/\l^2$, it is more convenient
first to collect similar terms in $G_\l^{(n)} v^\dagger $
and $v G_\l^{\dagger\,(n)}$ to the given order $c^n$.
After this each term in $G_\l^{(n)} v^\dagger $ is orthogonal to all terms
in $v G_\l^{\dagger\,(n)}$ except for one.
Whether or not this kind of an iterative procedure
could be useful for deriving recurrence relation between meander
numbers would depend on a possibility to
reexpand~\rf{As} in such orthogonal terms in the general case.

We see now the difference between the expansion in $v$ for the meander
problem, described in this Subsection,
and that for the two-matrix model with a polynomial potential~\cite{DL95}.
The latter reduces to an algebraic equation while the former does not.
This differs the matrix models describing the meander numbers
from the Kazakov--Migdal model whose solution can be described
in the language of the two-matrix model~\cite{DMS93}.

\newsection{Discussion}

We have considered in this paper the meander problem which results in more
complicated matrix models than those solved before. It belongs to the same
generic class of problems of words as, say, the large-$N$ QCD in $D=4$ but is
presumably simpler.

We have introduced a new representation of the meander numbers
via the general complex matrix model. This allowed us to consider the
supersymmetric matrix model
which includes both bosonic and fermionic matrices and
describes the principal meanders. This model looks simpler than a pure bosonic
model and one can try to solve it by using supersymmetry Ward identities.

Using non-commutative sources, we have reformulated
the meander problem as that of averaging in a Boltzmannian
Fock space whose annihilation and creation operators
obey the Cuntz algebra and have shown the equivalence with the
combinatorial approach based on the arch statistics. 
The averaging expression is represented
in the form of a product of two continued fractions but further progress
requires their better understanding as functions of the non-commutative
variables. One of the possible approaches can be based on the
standard procedure of disentangling via the path integral.


We have discussed also the relation between the matrix models
describing the meander problem and the Kazakov--Migdal model
on a $D$-dimensional lattice.
The words are the same both for the meander problem
and for the Kazakov--Migdal model while the only difference resides in the
meaning of nonvanishing words.
This relation could give a hint on how to solve the meander problem.

We have demonstrated how the solution of the Kazakov--Migdal model with
the quadratic potential can be obtained using the theorem
of addition of free random variables.
We have shown that this approach does not work for the meander problem
even within the interaction representation
since the variables are not free for this case
so that the theorem of addition is not applicable.
This observation casts doubts that a solution of large-$N$ QCD can be
obtained in the language of a masterfield given by free random variables.

The supersymmetric matrix models of the type discussed in this paper
for the meander problem are novel ones.
It is worth studying them in connection with other physical applications,
in particular, with a discretization of super-Riemann surfaces and
superstrings.

\subsection*{Acknowledgments}

We are grateful to J.~Ambj{\o}rn, L.~Chekhov, P.~Cvitanovi\'{c},
C.~Kristjansen, G.~Semenoff, K.~Zarembo, and J.-B.~Zuber
for useful discussions.
The research described in this publication was supported in part by
the International Science Foundation under Grant MF-7300.

\eop

\setcounter{section}{0}
\setcounter{subsection}{0}

\appendix{Derivation of \protect{\eq{numgraphs}}
via free random variables \label{ss:r.v.}}

Equation~\rf{numgraphs} can be derived using free random variables
 as follows. First of all let us note that
\bea
\lefteqn{\#_n=
\# \ {\rm of\ nonvanishing\ terms\ in\ } \LA \frac 1N \tr \LB
\sum_{a,b=1}^m W_a W_b^\dagger\RB^n\RAG} \non &= & \# \ {\rm \
of\ nonvanishing\ terms\ in\ } \LA \frac 1N \tr \LB \sum_{i=1}^m
 A_i\RB^{2n}\RAG
\eea
where the matrices $A_i$ are Hermitean.
The simplest nontrivial example is that of $m=2$, when the LHS has for $n=2$
six nonvanishing terms of the type of
$W_1 W_1^\dagger W_1 W_1^\dagger$, $W_1 W_2^\dagger W_2 W_1^\dagger$,
$W_1 W_1^\dagger W_2 W_2^\dagger$,
$W_2 W_2^\dagger W_2 W_2^\dagger$, $W_2 W_1^\dagger W_1
W_2^\dagger$, $W_2 W_2^\dagger W_1 W_1^\dagger$.
The RHS has six nonvanishing terms of the type of $A^4$,
$A^2B^2$, $AB^2A$, $B^4$, $B^2A^2$, $BA^2B$, where we have used the
notations $A_1\equiv A$, $A_2\equiv B$ --- the same as in the Table~3.

If we now introduce free random variables $\widehat A_i$, such  that
\bea
\LA \Omega \left| \widehat A_i^{2n}\right| \Omega \RA &=&1 \,, \non
\LA \Omega \left| \widehat A_i^{2n+1}\right| \Omega \RA& =&0\,,
\eea
then
\be
\LA \Omega \left| \LB\sum_{i=1}^m \widehat A_i \RB^{2n} \right| \Omega \RA =
\#_n
\ee
since the vacuum average of each nonvanishing term equals one.
The resolvent for each $\widehat A_i$ reads
\be
R_i(z)
\equiv  \sum _{n=0}^\infty
\frac{\LA \Omega \left| \widehat A^{n}_i\right| \Omega \RA }{z^{n+1}}
=\sum _{n=0}^\infty \frac{1}{z^{2n+1}} = \frac{z}{z^2-1} \,.
\label{1resolvent}
\ee

The inverse function to~\rf{1resolvent} is given by
\be
A_i(z)\equiv R_i^{-1}(z)=\frac{\sqrt{1+4z^2}+1}{2z} \,.
\ee
Then the master-field operator can explicitly be constructed
as~\cite{Dou95,GG95}
\be
\widehat A_i (a,a^\dagger)= a+\frac{\sqrt{1+4(a^\dagger)^2}-1}{2a^\dagger}
= a+a^\dagger - (a^\dagger)^3 +\cdots \,.
\ee
where the RHS is understood as the expansion around zero.

The theorem of addition of free random variables states that if
$\widehat A_i$'s are
free random variables and so they are representable as
\be
\widehat A_i=a_i+ f_i(a^\dagger _i)\,,
\ee
 where $f_i$'s are functions analytic around zero, then
\be
\LA \Omega \left| \LB \sum_i \widehat
A_i \RB^n \right |\Omega \RA = \LA \Omega \left| \LB a+\sum_i f_i(a^\dagger
) \RB^n \right| \Omega \RA \,.
\ee
Using this theorem, we get for the $m$
matrices
\be
A(z) = \frac 1z + m\left(A_i(z) -\frac 1z \right) =
\frac{m\sqrt{1+4z^2}+2-m}{2z} \,.
\ee
The resolvent is given by the inverse function
\be
R(z)= A^{-1}(z) = \frac{(2-m)z + m \sqrt{z^2-4(m-1)}}
{2\left(z^2-m^2 \right)} \,.
\label{mresolvent}
\ee
Equation~\rf{numgraphs} can be reproduced substituting $z=1/c$
and dividing by $c$.

As far as the Kazakov--Migdal model is concerned,
\eq{defrM} results in the following representation
of the correlator~\rf{KMcorrelator} via the words built up of
the $m=2D$ unitary matrices:
\bea
\LA \frac 1N \tr{\phi^2(0)} \RA = \sum_{n=0}^{\infty }\; c^{2n}
\sum_{a_1,a_2,\cdots, a_{2n}=1}^{2D}
\LA \frac 1N \tr
\left( U_{a_1} U_{a_2}^\dagger \cdots
U_{a_{2n-1}} U_{a_{2n}}^\dagger\right) \RAH^2 \,.
\eea
Since the meaning of each word is either one or zero,
the square on the RHS is not essential.
Equation~\rf{numgraphs} with $m=2D$ then yields
\be
\LA \frac 1N \tr{\phi^2(0)} \RA
= \frac {D\sqrt{1-4(2D-1)c^2}-D+1}{1-4D^2c^2}
\ee
which reproduces the solution of Ref.~\cite{Gro92}.

We discuss in the Appendix~C that the approach, which is described here for
the Kazakov--Migdal model, does not work for the meander problem
since the random variables are not free in the latter case.

\appendix{Combinatorial interpretation of the results of Sect.~3}

The representation of Sect.~3 of the meanders via non-commutative random
variables can be alternatively derived pure combinatorially which
clarifies the relation between our approach and that of Ref.~\cite{DGG95}.

Equation~\rf{SD} can be derived combinatorially as follows.
Let us denote by ${\cal
G}_n$ the sum of all possible arch configurations of order $n$ with $m$
colorings. For example, ${\cal G}_1$ for $m=2$ is given by

\mbox{}

\be
{\cal G}_1=~~~~~~~~~~~~~~~~~~~~~~~~~~~~~\,,
\ee

\begin{picture}(150,10)(-5,4)
\put(180,30){\oval(30,30)[t]}
\put(210,30){$+$}
\thicklines\put(250,30){\oval(30,30)[t]}
\end{picture}
\vspace*{-.5cm} \mbox{}

\noindent
where the two terms on the RHS differ by the coloring.
Picking up the leftmost arch, we have
the following recurrence relation for ${\cal G}_n$'s:
\be
{\cal G}_n=\sum_{k=0}^{n-1}\sum_{{\rm colors}\atop {\rm of}~\cap}~~~~~
{\cal G}_k~~~~~
{\cal G}_{n-k-1}\,,~~~~~~~~~ {\cal G}_0=\emptyset \,.
\label{rerel}
\ee
\begin{picture}(150,10)(2,7)
\put(205,50){\oval(30,30)[t]}
\end{picture}
\vspace*{-.7cm} \mbox{}

Let us analogously
denote by ${\cal G}^\dagger_n$ the sum of all possible upside-down
arch configurations of order $n$ with $m$ colorings. For example, ${\cal
G}_1^\dagger$ for $m=2$ is given by

\be
{\cal G}^\dagger_1=~~~~~~~~~~~~~~~~~~~~~~~~~~~~~~~~.~~~~~~~
\ee
\begin{picture}(150,10)(-5,8)
\put(180,40){\oval(30,30)[b]}
\put(210,30){$+$}
\thicklines\put(250,40){\oval(30,30)[b]}
\end{picture}
\vspace*{-.3cm} \mbox{}

\noindent
${\cal G}^\dagger_n$ obeys the recurrence relation which is similar to
\eq{rerel}.

Let us now introduce the ``multiplication'' $\circ$ of arch
configurations. Its meaning is evident from the following examples:\par
\begin{picture}(100,80)
\put(100,50){\oval(20,20)[t]}
\put(125,50){\oval(20,20)[t]}
\put(150,50){$\circ$}
\put(190,50){\oval(40,40)[b]}
\put(190,50){\oval(20,20)[b]}
\put(220,50){$=$}
\put(260,50){\oval(40,40)[b]}
\put(260,50){\oval(10,10)[b]}
\put(247.50,50){\oval(15,15)[t]}
\put(272.50,50){\oval(15,15)[t]}
\put(290,50){,}
\put(125,10){\oval(20,20)[t]}
\put(150,10){$\circ$}
\thicklines\put(180,10){\oval(20,20)[b]}
\put(220,10){$=$}
\put(240,10){$\emptyset$}
\put(250,10){.}
\end{picture}

We have then the relation
\be
{\cal G}_n\circ {\cal G}_n^\dagger =\left\{{\rm the\ sum\ of\ all\ possible}
\atop \hbox{multi-component\ meanders}\right\}
\,,
\ee
and
\be
\#\ {\rm of\ terms\ in\ }{\cal G}_n\circ {\cal G}_n^\dagger=\sum_{k=1}^n
M_n^{(k)}m^k
\,,
\label{numofterm}
\ee
since each loop in the meander can be colored with $m$ colors.

The ``multiplication'' $\circ$ can naturally be represented in the
operator language as follows. Let us
introduce $m$ non-commuting  variables  $\u_a$ ($a=1,\ldots ,m$), one for each
color in the arch configurations.  Let us associate with each arch
 configuration a word made up of $\u_a$'s in a way, which is evident from the
following   example:\par
\begin{picture}(100,60)
\put(60,40){\oval(20,20)[t]}
\thicklines\put(90,40){\oval(20,20)[t]}
\put(120,40){$\longleftrightarrow~~~ \u_1\u_1\u_2\u_2,$~~~~if~~~~$\u_1
\leftrightarrow~~~~~~~~$and~~$\u_2 \leftrightarrow$~~~~~~~.}
\put(370,42){\line(1,0){15}}
\thinlines\put(285,42){\line(1,0){15}}
\end{picture}
\vspace*{-.9cm} \mbox{}

\noindent
Analogously, we associate words, made up of  $\u_a^\dagger$'s, with the
upside-down arch configurations as in the example \par
\begin{picture}(100,60)
\put(30,45){\oval(20,20)[b]}
\thicklines\put(60,45){\oval(20,20)[b]}
\put(90,40){$\longleftrightarrow~~~ (\u_1\u_1\u_2\u_2)^\dagger=\ud_2\ud_2
\ud_1\ud_1,
$~~~~if~~~~$\u_1
\leftrightarrow~~~~~~$and~~$\u_2 \leftrightarrow$~~~~~~~.}
\thinlines\put(333,42){\line(1,0){15}}
\thicklines\put(413,42){\line(1,0){15}}
\end{picture}
\vspace*{-.9cm} \mbox{}

We have from~\eq{rerel} the following equation in terms of $\u_a$'s:
\be
G_n(\u)=\sum_{k=0}^{n-1}\sum_{a=1}^m \u_a G_k(\u) \u_a G_{n-k-1}(\u)
\,,
\label{urerel}
\ee
where $G_n(\u)$ stands for the sum of the words of order $n$.
Introducing the generating function (cf.~\rf{defG})
\be
G_\l (\u)=\sum_{n=0}^{\infty}G_n(\u)\l^{-2n-1} \,,
\label{dfGl}
\ee
\eq{urerel} can be rewritten in the form~\rf{SD}.

There is the correspondence
\be
{\cal A}\circ {\cal B}\longleftrightarrow
\LA\Omega\left|A(\u)B(\ud)\right|\Omega\RA \,.
\ee
Using this correspondence, Eqs.~\rf{numofterm} and \rf{dfGl}, we arrive at
\eq{nondefM}.

Equation~\rf{SSD} can be interpreted in the same way.
As an example, let us consider the
supersymmetric case of $m=2$, when $\u_a=(u,v)$.
The analog of~\eq{rerel} for this case reads
\bea
{\cal G}_0&=&\emptyset \,, \non
{\cal G}_n&=&\sum_{k=0}^{n-1}\left[~~~~\bar{{\cal G}}_k~~~~{\cal G}_{n-k-1}+
~~~~\bar{{\cal G}}_k~~~~{\cal G}_{n-k-1}\right]
\,,
\eea
\begin{picture}(150,20)
\put(189,45){\oval(30,30)[t]}
\thicklines\put(282,45){\oval(30,30)[t]}
\end{picture}
\vspace*{-.9cm} \mbox{}
\par \noindent
where
\bea
{\cal G}_n&=&{\cal G}_n(\cap ,\bbox{\cap})\,, \non
\bar{{\cal G}}_n&=&{\cal G}_n(\cap ,-\bbox{\cap})\,.
\eea

Let us introduce the projector ${\cal P}$, such that it picks up the arch
configurations whose rightmost arch is of light color. For example,\par
\begin{picture}(100,30)(0,0)
\put(40,5){${\cal P}$}
\put(50,5){$\left[~~~~~~~~~+~~~~~~~~~~~~\right]=~~~~~~~~.$}
\put(70,5){\oval(15,15)[t]}
\put(120,5){\oval(15,15)[t]}
\thicklines\put(120,5){\oval(30,30)[t]}
\thinlines\put(180,5){\oval(15,15)[t]}
\end{picture}
\par \noindent
Then, we have
\be
\left\{ {\rm the\ sum\ of\ all\ principle\atop meanders\ of\ order\
n}\right\}={\cal P}\left[{\cal G}_n\right]\circ{\cal P}\left[\bar{{\cal
G}}^\dagger_n\right] \,.
\ee
The
projector ${\cal P}$ corresponds in the operator language to the insertion
of $u^\dagger u$ in the \eq{barsmea}.

\appendix{Comment on freeness of random variables for the meander problem}

We demonstrated in the Appendix~A how the Kazakov--Migdal model
can be solved via free random variables.
We comment here that this approach does not work for the meander
problem since the random variables are not free for this case.

The meander numbers can be represented by
the vacuum average of certain non-commuting variables as
\be
\sum_{k=1}^n M_n^{(k)} m^k = \frac 1{N^2} \LA  {\rm Tr} \left ( \sum_{i=1}^m
 A_i\otimes A_i \right )^n \RAG = \LA \Omega  \left | \left
(\sum_{i=1}^m \widehat A_i\widehat {\tilde A}_i \right )^n \right | \Omega
\RA \,,
\label{C1}
\ee
where the operators $\widehat A_i$'s
and $\widehat {\tilde A_i}$'s under the vacuum averaging are
\be
\widehat{A}_i= a_i+ a_i^{\dagger}\, , ~~~~
\widehat {\tilde A}_i= \tilde a_i + \tilde a_i^{\dagger}\,.
\ee
Here the two sets of operators $a_i$'s and $\tilde a_i$'s
obey the Cuntz's algebra
\be
a_ia^\dagger_j =  \delta_{ij}\,, ~~~
\tilde a_i\tilde a^\dagger_j = \delta_{ij}
\ee
independently of each other while all the operators with tildes commute
with all the operators without tildes.
The equivalence of the two representations in \eq{C1}
is evident from the factorization
\be
\LA  {\rm Tr} ( A \otimes B) \RA \equiv \LA \tr A \tr B \RA
= \LA \tr A \RA \LA \tr B\RA
\ee
at large $N$.

Equation~\rf{C1} involves the direct product of matrices
which is represented as the product of two independent operators
$\widehat {A}_i$ and $\widehat {\tilde A}_i$.
The non-commuting variables of the type
$\widehat {A}_i\widehat {\tilde A}_i$ are not
free random variables of
Ref.~\cite{VDN92} since they do not satisfy the defining axiom of
free random variables. This can be seen by considering the average
\bea
\lefteqn{\LA \Omega \left | \left [(\widehat {A}\widehat {\tilde
A})^2-\LA (\widehat {A}\widehat {\tilde A})^2\RA \right ]
\left [ (\widehat {B}\widehat {\tilde B})^2 -
\LA (\widehat {B}\widehat {\tilde B})^2\RA
\right ] \right.\right. }  \non & &\times \left.\left.
\left [ (\widehat {A}\widehat {\tilde A})^2
-\LA (\widehat {A}\widehat {\tilde A})^2\RA \right ]
\left [ (\widehat {B}\widehat {\tilde B})^2
-\LA (\widehat {B}\widehat {\tilde B})^2\RA \right ] \right | \Omega \RA
\stackrel{?}= 0
\label{C4}
\eea
which would vanish if they were free random variables.

It is easy to see by direct calculation that the expression~\rf{C4}
does not vanish. We rewrite it as
\bea
\lefteqn{
\LA \Omega \left| \left[ (\widehat {A}\widehat {\tilde A})^2-1 \right]
\left[ (\widehat {B}\widehat {\tilde B})^2 - 1
\right] \left[ (\widehat {A}\widehat {\tilde A})^2 -1 \right]
 \left[ (\widehat {B}\widehat {\tilde B})^2- 1 \right]
\right| \Omega  \RA} \non & &=
\LA \Omega  \left|  \widehat {A}^2\widehat B^2
\widehat A^2\widehat B^2  \right| \Omega  \RA^2
-\LA \Omega  \left|  \widehat A^4   \right| \Omega  \RA^2
-\LA \Omega  \left|  \widehat B^4   \right| \Omega  \RA^2
+1 ~~~~~~~~~~~~~ \non & &
=3^2-2^2-2^2 +1=2
\label{notfree}
\eea
since
\be
\LA \Omega  \left|  (\widehat A \widehat{ \tilde A})^2  \right| \Omega  \RA =
\LA \Omega  \left|  (\widehat B \widehat{ \tilde B})^2  \right| \Omega  \RA = 1 \,.
\ee
The squares on the RHS of \eq{notfree} are due to analogous contributions
from the operators with tildes.

The nonvanishing RHS of \eq{notfree} means that the variables
$\widehat{ A}_i\widehat{\tilde A}_i$
are {\it not\/} free. This is, as is already mentioned, a consequence of
the direct product of matrices which results in the product of
$\widehat{A}_i$ and $\widehat{\tilde A}_i$
 and leads, in turn, to the squares
in \eq{notfree}. If there are no squares, the axiom of freeness is
obviously satisfied. The same is true for the supersymmetric case as well.

The example of this Appendix shows that random variables are not necessarily
free in (multi-) matrix models with complicated interaction even within
the framework of the interaction representation.
The simplest average for the meander problem which violates the axiom of
freeness is that~\rf{C4}. For this reason, the theorem of addition with
\be
R_i(z) \equiv  \sum _{n=0}^\infty
\frac{\LA \Omega \left| (\widehat A_i \widehat {\tilde A}_i)^n
\right| \Omega \RA }{z^{n+1}}
=\sum _{n=0}^\infty \frac{C^2_n}{z^{2n+1}} \,,
\label{1meresolvent}
\ee
where $C_n$ are given by \eq{Catalan},
can be used, quite similarly to the Appendix~A, to describe the meander numbers
only up to $n=3$ but fails to reproduce $M_n$ for $n\geq4$.

\eop

\end{document}